\documentclass{ws-rv1025x75}
\usepackage{ws-rv-van}             % numbered citation/references (default)
\usepackage{mathrsfs}
\usepackage{amsmath, amsfonts, amssymb}
\usepackage{graphicx}
\usepackage{bm}
\usepackage{color}
\usepackage{amsfonts}

\usepackage{amscd}
\newcommand{\bea}{\begin{eqnarray}}
\newcommand{\eea}{\end{eqnarray}}

\def\beq{\begin{equation}}
\def\eeq{\end{equation}}

\def\pno{\par\noindent}

\def\t0{\tau_{_0}}

\makeindex
\begin{document}

\chapter[Mesoscopic Aharonov-Bohm Interferometers: Decoherence, Persistent Currents and Thermoelectric Transport] {Mesoscopic Aharonov-Bohm Interferometers: \\ Decoherence, Persistent Currents and Thermoelectric Transport\label{ch1}}
%\footnotetext{Title footnote.}
\author[O. Entin-Wohlman, A. Aharony, and Y. Imry]{O. Entin-Wohlman$^{1,2}$, A. Aharony$^{1,2}$,  and Y. Imry$^3$}
%\index[aindx]{Author, F.} % or \aindx{Author, F.}
%\index[aindx]{Author, S.} % or \aindx{Author, S.}

\address{$^1$Raymond and Beverly Sackler School of Physics and Astronomy, Tel Aviv University, \\Tel Aviv 69978, Israel\\
$^2$Department of Physics and the Ilse Katz Center for Meso- and Nano-Scale Science\\
and Technology, Ben Gurion University, Beer Sheva 84105, Israel\\
$^3$Department of Condensed Matter Physics,  Weizmann Institute of Science, \\Rehovot 76100, Israel\\
oraentin@bgu.ac.il, aaharony@bgu.ac.il, yoseph.imry@weizmann.ac.il}

\begin{abstract}
Two important features of mesoscopic Aharonov-Bohm (A-B) electronic interferometers are analyzed: decoherence due to coupling with other degrees of freedom and the coupled transport of charge and heat. 
We first  review the principles of decoherence of electronic interference. 
We then analyze the thermoelectric transport in a ring threaded by such a flux, with a molecular bridge on one of its arms. The charge carriers may also interact  inelastically with the molecular vibrations.  This nano-system is connected to three terminals; two of them are electric and thermal,  held at slightly different chemical potentials and temperatures, and the third is purely thermal. For example, a phonon bath thermalizing the molecular vibrations.   When this third terminal is held at a temperature different from those of the electronic reservoirs, both an electrical and a heat current are, in general, generated between the latter.
Likewise, a voltage and/or temperature difference between the electronic terminals leads to thermal current between the thermal and electronic terminals. The transport coefficients governing these conversions (due to energy exchange between the electrons and the vibrations) and their dependences on the A-B flux are analyzed. 
%and shown to obey the full Onsager-Casimir relations. 
Finally, the decoherence due to these inelastic events is discussed.
\end{abstract}

\body

\section{Introduction}\label{sec1}
Interference, resulting from the superposition of different amplitudes, is a basic attribute of Quantum Mechanics.
Akira Tonomura Sensei has made decisive lasting contributions to the study of these effects for electron beams. \cite{Akira}
Here we shall review the analogous Physics in mesoscopic solid-state systems\cite{book}. Here, small sizes 
and/or low temperatures are necessary in order to keep electrons coherent. Clearly, the understanding
of the processes of electronic decoherence is crucial here.
We shall consider an Aharonov-Bohm
(A-B) interference experiment on a  ring-type structure. 
This proves to be a
convenient way to observe interference patterns in such samples, providing an experimentally straightforward way of shifting the interference pattern.
This experiment involves two electron wave packets, $\ell(x)$ and
$r(x)$ ($\ell,r$ stand for left, right), crossing the ring along its two
opposite sides and interacting with an environment, represented by a (usualy thermal) bath.
%(see Fig. ~\ref{sys}, 
%where the
%left and right paths are the upper and lower ones and the thermal reservoir is the environment)
We assume that the two wave-packets follow  classical
paths, $x_\ell(t), x_r(t)$ along the arms of the ring. The interference is examined
after each of the two wave packets had traversed half of the ring's circumference.
The combined wave function is then
\begin{align} \psi=\ell(x)+r(x)\ .
\label{int}  \end{align}
\noindent The mixed terms, $2 {\rm Re}[\ell(x)\times r^*(x)]$, constitute the interference contribution,
clearly sensitive to a phase shift (e.g. AB) introduced between the two partial waves $\ell(x)$ and $r(x)$.
Interaction with the environment can reduce the strength of the interference. This process is called  ``decoherence"
and it is of great relevance in this paper.

We will review the effects of interference, first on the purely electronic transport.
Then, on coupled 
electric and thermal transport. When changes in the environment state, caused by the electrons,  occur, decoherence may follow. These changes will be modeled by the effect of a local vibrational mode (``vibronic coupling"), which is in turn coupled to a reservoir that is restoring equilibrium to the local vibration.

Thermoelectric effects in bulk conductors usually necessitate
breaking of particle-hole symmetry, which can be  
substantial and controllable in mesoscopic structures. As a result there is currently
much interest  in  investigations of thermoelectric
phenomena   in nanoscale devices   at low temperatures.
%Experimental studies have been carried out on point-contacts,
%\cite{PELTIER,MOLENKAMP}  quantum dots and semiconducting systems , \cite{QDTP,PEPPER,LUDOPH}
%nanotubes, \cite{COHEN,PEREZ,MCEUEN} silicon nanowires,
%\cite{HOCHBAUM} and more.

 Very recently, the paramount importance of another symmetry breaking has been pointed out.
It has been  proposed that the thermal efficiency of a thermoelectric heat engine, or refrigerator, can be significantly enhanced once {\it time-reversal
symmetry is broken} (by, e.g., a magnetic flux). The Onsager
symmetry of  the thermopower coefficients is accordingly modified. \cite{benenti} 

%Early theoretical studies of
%thermoelectric  transport coefficients of microstructures were
%based on the Landauer approach
%\cite{SIVAN,STREDA,BUTCHER,PROETTO,CBEENAKKER} and were mainly
%focused on charge and heat currents between two electronic
%terminals, without coupling to phonons. Later on, effects of
%electron-electron processes and electronic correlations
%(increasingly important at lower temperatures), as well as that of
%an applied magnetic field, on the thermopower produced in large
%\cite{MATVEEV} and single-level \cite{KIM} quantum dots,  and also
%in  quantum wires \cite{flensberg} were considered. The signature
%of attractive electronic interactions on the thermopower was
%considered in Ref. ~\onlinecite{Karen}, and the dependence of the
%thermoelectric response on the length of the atomic chain
%connecting the leads has been recently computed within a
%density-functional theory. \cite{CUEVAS}

The coupling of the charge carriers to vibrational modes of
the molecule should play a significant role in thermoelectric
transport through molecular bridges. %, even more so in the nonlinear
%regime. \cite{FLEN} Indeed, a density-functional computation of
%the nonlinear differential conductance of gold wires attributed
%changes in the I-V characteristics to phonon heating,
%\cite{FREDERIKSEN,NITZAN} and the 
The thermopower coefficient was
proposed as a tool to monitor the excitation spectrum of a
molecule forming the junction between two leads, \cite{ERAN,FINCH}
and
to determine both the location of  Fermi level of the charge carriers 
and their charge. \cite{MURPHY}.%, either for a molecular conductor,
%\cite{PAULSSON,TAIWAN,baranger} or for an atomic chain.
%\cite{ZHENG,DVIRA} 

Theoretically, when the coupling to the vibrational modes is
ignored, the transport coefficients are determined  by the usual energy-dependent transmission
coefficient,  replacing  the conductivity.
\cite{CM,SIVAN,CUEVAS} Even when the corrections to the thermoelectric
transport due to the coupling to the vibrational modes are 
small, their study is  of interest because of fundamental
questions related to the symmetries of the conventional transport
coefficients, and since they give rise to additional coefficients
connecting the heat transport in-between the electrons and the
vibrational modes. Recently, \cite{NEWWE} %(referred to
%below as I) 
we have analyzed these phenomena in a
molecular bridge. 
In particular we have considered the case where the molecule
is (relatively) strongly coupled to a heat bath of its own, which maintains  its temperature different from those of the source and sink of
the charge carriers.    Namely, we have  assumed
that the relaxation time due to the coupling of the molecule to
its own heat bath, $\tau^{}_{\rm V}$, is short on the scale of the
coupling of the molecular vibrations to the charge carriers. This coupling, $\gamma$  is our
snall parameter.  $\hbar/ \tau^{}_{\rm V}$ may still
be very small on all other physical scales, such as
$\hbar\omega^{}_0$, where $\omega^{}_0$ is the frequency of the
vibration, or the molecular (electronic) level width. The phonon
bath may be realized  by an electronically insulating substrate or
a piece of such material touching the junction, each held at  a
temperature $T^{}_{\rm V}$. %%consisting  of quantum dots, were discussed in  Refs. ~\onlinecite{EDWARDS},  ~\onlinecite{SL},
%and ~\onlinecite{CAMB}.   
%The effects of thermal probes  on the electronic heat
%conduction have been considered theoretically in some cases.
 %\cite{Hekking,DAVID} 
 Experimental realizations of  three-terminal setups have been discussed. \cite{EDWARDS,SL,CAMB} %n  Refs. ~\onlinecite{EDWARDS},  ~\onlinecite{SL},
%and ~\onlinecite{CAMB}.   

\begin{figure}[ hbtp]
\begin{center}
\includegraphics[width=6cm]{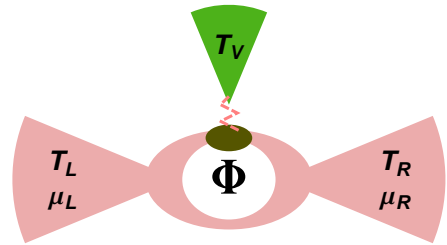}
\caption{A mixed, thermal-electronic, three-terminal system,
as explained in the text, threaded by a magnetic flux
$\Phi$. }\label{Fig1}
\end{center}
\end{figure}

We organize this paper as follows.
Section 2  discusses the principles of decoherence.  Our  theoretical model  is presented in section 3 and the A-B oscillations of the transmission of the interferometer and their decoherence are discussed in section 4.
After a short review   of thermoelectricity   in section 5, we give the  main results of the analysis including the magnetic-flux dependence of the three-terminal transport coefficients
Section 6 presents these coefficients of the three-terminal junction in the linear-response regime and verifies their Onsager symmetries.  Section 7 is devoted to a discussion of the results  in  simple examples.  Section 8 summarizes the article.

\section{Principles of decoherence}\label{sec2}
Interference effects are strongly
affected by the coupling of the interfering particle to its environment, e.g., to a
heat bath.  The way such a coupling modifies quantum phenomena has been studied for a
long time, both theoretically,  \cite{FV,CL}  %(Feynman and Vernon \cite{FV}, Caldeira and Leggett \cite{CL}),
and experimentally.  \cite{Buks}  Some of the effect of the coupling to the environment may be
described by the ``phase-breaking" time, $\tau_\phi$, which is the
characteristic time for the interfering particle to stay phase coherent as explained
below.
\par   Our discussion will be based on the work of Stern {\it et al.} \cite{Stern}
Two points of view have been used to describe how the interaction of a quantum system
with its environment might suppress quantum interference. The first regards the
environment as measuring the path of the interfering particle. When the environment
has all the information on that path, no interference is seen. The second description
considers the phase uncertainty induced on the interfering particle by the interaction with the environment.  The two descriptions were proven to be equivalent. Here we will review the first
point of view mentioned above, using
the example  of an A-B %Aharonov-Bohm
%(A-B) 
interference experiment on a ring. The electron (whose coordinate is $x$) and
the environment (whose wave functions for the two paths are denoted by
$\chi$ and $\eta$) interact during the traversal of the ring. As a result, the combined wave function of Eq. \ref{int}
becomes
\begin{align} \psi=\big[\ell(x)+r(x)\big ]\otimes\chi_{_0}(\eta)
\label{3.1}  \end{align}
\pno
At time $\tau_{_0}$ (measured with respect to the start of the experiment), when
the interference is examined, the wave function is, in general,
\begin{align} \psi(\tau_{_0})=l(x,\tau_{_0})\otimes\chi_{_\ell}(\eta,\t0)\ +\
r(x,\tau_{_0})\otimes\chi_{_r}(\eta,\t0)\label{3.2} \end{align}
%\eqno(\mbox{III.2})
and the interference term is (see below),
\begin{align} 2\ {\rm Re}\ \Big [\ell^*(x,\tau_{_0})r(x,\tau_{_0})\int\!
d\eta\chi_{_\ell}^*(\eta,\t0)\chi_{_r}(\eta,\t0)\Big ]\label{3.3}\ .\end{align}
Without the coupling to the  environment t, the interference term would
have been just Eq. \ref{int}. So, the effect of the
interaction is to multiply the interference term by the scalar product $\int\! d\eta\chi_{_\ell}^*(\eta,\t0)
\chi_{_r}(\eta,\t0)$.  The first way to
understand the dephasing is clear from this expression, i.e. the reduction in the interference due to the response of the  environment to the interfering waves.  When the two states of the environment become
orthogonal, the final state of the environment identifies the path the electron took.
Quantum interference, which is the result of an uncertainty in this path, is then
lost. The phase breaking time, $\tau_{_\phi}$, is the time at which the two
interfering partial waves shift the environment into states to become orthogonal to each other,
i.e., when the environment has the full information on the electron path.
\par As seen from the above discussion, the phase uncertainty remains constant when the
interfering wave does not interact with the environment. Thus, if a trace is left by
a partial wave on its environment, this trace cannot be wiped out after the
interaction is over.  This statement can be
proven also from the point of view of the change the interfering wave induces in its
environment. This proof follows simply from unitarity. The scalar product of two
states evolving under the same Hamiltonian does not change in time.
%Therefore, if the state of the system (electron plus environment) after the
%electron-environment interaction took place is
%\begin{align}
% |r(t)\ra\otimes |\chi^{}_{\ell}%\chi_{_{env}}^{(1)}
% \ra+ |\ell (t)\ra\otimes
%|\chi^{}_{r}%\chi_{_{env}}^{(2)}
%\ra\ ,\label{3.20}
%\end{align}
%then the scalar product $\la\chi^{}_{\ell}%\chi_{_{env}}^{(1)}
%(t)|\chi^{}_{r}%\chi_{_{env}}^{(2)}
%(t)\ra$ does not change with time. 
The only way to change it is by another interaction of the
electron with the same environment (see the discussion at the end of this section). 
%Such
%an interaction keeps the product
%$\la\chi^{}_{\ell}%\chi_{_{env}}^{(1)}
%(t)|\chi^{}_{r}%\chi_{_{env}}^{(2)}
%(t)\ra\otimes\la r(t)|l(t)\ra$ constant,
%but changes $\la\chi^{}_{\ell}%\chi_{_{env}}^{(1)}
%(t)|\chi^{}_{r}%\chi_{_{env}}^{(2)}(t)\ra$.    
The interference
will  be retrieved only if the orthogonality is transferred from the environment wave
function to the electronic wave functions which are not traced on in the experiment.
\par The above discussion was concerned with the phase $\phi = \phi_r$, accumulated by
the right-hand path only. The left hand path accumulates similarly a phase
$\phi_{_\ell}$ from the interaction with the environment. The interference pattern is
governed by the {\it relative} phase $\phi_r-\phi_{_\ell}$, and it is the uncertainty
in {\it that} phase which determines the loss of quantum interference. This
uncertainty is always smaller than, or equal to, the sum of uncertainties in the two
partial waves' phases. 
\par Often the same environment interacts with the two interfering waves. A typical
example is the interaction of an interfering electron with the electromagnetic
fluctuations in vacuum. In this case, if the two waves follow parallel paths with
equal velocities, their dipole radiation, despite the energy it transfers to the
field, does not dephase the interference. This radiation makes each of the partial
waves' phases uncertain, but does not alter the relative phase. Another well-known 
example is that of  ``coherent inelastic neutron scattering" in crystals. \cite{Kit} 
This process follows from the coherent addition of the
amplitudes for the processes in which the neutron exchanges {\it the same} phonon with
{\it all} scatterers in the crystal.
\par The last example also demonstrates that an exchange of energy is not a sufficient
condition for dephasing. It is also not a necessary condition for dephasing. What
is important is that the two partial waves flip the environment to {\it orthogonal}
states. It does not matter in principle that these states are degenerate. Simple
examples were given by Stern {\it et al.} \cite{Stern}  Thus, it must be emphasized that, for
example, long-wave excitations (phonons, photons) cannot dephase the interference.
But that is {\it not} because of their low energy but rather because they do not
influence the {\it relative} phase of the paths.
\par We emphasize that dephasing may occur by coupling to a discrete or a continuous
environment. In the former case the interfering particle is more likely to ``reabsorb"
the excitation and  ``reset" the phase. In the latter case, the
excitation {\it may} move away to infinity and the loss of phase can then be broadly regarded
as irreversible. The latter case is that of an effective ``bath". \cite{book}
%and there are no subtleties with the definition of the relative phase, $\phi$, %between the two paths. 

We point out that in special cases it is possible, even in the
continuum case, to have a finite probability to reabsorb the
created excitation and thus retain coherence. This happens, for
example, in a quantum-interference model due to Holstein \cite{ho}
for the Hall effect in insulators (see also Refs. 
~\cite {Mike} and   %Entin-Wohlman et al ~ 
~\cite{E-W}). This model deals with
phonon-induced hopping between two localized states with different
energies, $E_1$ and $E_2$, which necessitates the (real) absorption of a 
phonon with energy $E_2 - E_1$ for $E_2 - E_1 > 0$. It focuses on transitions which occur
via an intermediate localized state with yet another energy,
$E_3$. We take, for definiteness $E_3 > E_1$. That intermediate
state transition also involves  the virtual absorption of a a
phonon with frequency $\omega_q$, which is then re-emitted with
the transition from $3$ to $2$. This higher-order process has a
perturbation theory energy-denominator which at ``resonance"
($\hbar \omega_q = E_3 - E_1$) produces a term
 $i \delta(\hbar \omega_q - E_3 + E_1)$. The integration over the
 continuous $\omega_q$ then yields a term with a $\pi/2$ phase
 shift in the transition amplitude between $1$ and $2$.

 References ~\cite {ho} and ~\cite{E-W} analyzed the effects of a
 superposition (or interference) of the above amplitude with the direct one between
 $1$ and $2$, possibly including a phase shift  $ \phi=2\pi\Phi/\Phi^{}_{0}$,
(where $\Phi_{0}=hc/e$  is the flux quantum)%(the magnetic flux measured din units of the flux quantum $\Phi_{0}=hc/e$)  %\phi$  
 between the
 direct and the compound transition paths, due {\it e.g.} to a magnetic  flux $\Phi$ enclosed between them,

 % $\Phi_{0}=hc/e$) % $\Phi = \phi \hbar c /e$ 
 %enclosed between these paths.
 Note that these two interfering paths involve  for $E_2 - E_1 > 0$ the absorption
 of {\em the same} phonon and they therefore stay coherent (see discussion above).
 Interestingly, it turns out that although this process creates
 a $\sin(\phi)$ term in some transition probabilities,
 it retains the basic Onsager symmetry in $\Phi$ of the 
conductivity, 
 \begin{align}
 \sigma_{xx} (\Phi) = \sigma_{xx} (-\Phi)\ .
 \end{align}
 and also 
%However, concerning the symmetry of 
the ``non diagonal"
 Hall conductivity, $\sigma_{xy}$,
 \begin{align}
 \sigma_{xy} (\Phi) = \sigma_{yx} (-\Phi)\ .
 \end{align}
%it holds as well. 
Since a $\sin(\phi)$ dependence is
allowed here by symmetry, there is no reason why it should not exist.
Such a dependence is generated, in fact, by the Holstein process
\cite{ho,E-W}. This is the basic reason why the Holstein process
furnishes an explanation for the Hall effect in (localized)
insulators.

The adaptation of the Holstein idea to transitions between two leads with continuous 
energy spectra is actually simpler than the original model, in which an additional phonon
is needed to conserve energy. Here an elastic, purely electronic, transition can be made from a 
full state on one side to an empty state on the other. This transition can again be 
accomplished via a superposition of the direct transition %, $t_{0}$, 
with a compound one through the molecule, involving again a ``resonant" virtual phonon. Clearly, here this  
effect is of order $\gamma^2$ (see below, $\gamma$ is the electron-phonon coupling). The calculations
to that order were done in Ref. ~\cite{Oramnon}. Again, there is no $\sin(\phi)$ term in either the 
electrical or the thermal conductivity. It exists, however,  in other ``non diagonal" transport coefficients, such as
the thermopowers \cite{Oramnon}, as will be
discussed in section 7.

%The calculations were done in powers of the electron-phonon
%coupling~$\gamma$ (see section 4), and were carried to order
%$\gamma^2$. The basic Holstein contribution is of this order, and
%so are the mixed term in the transition probabilities. Thus, this
%order is sufficient for our purposes.

\section{The  model}\label{sec4}
In our analysis, the molecular bridge is represented by a single
localized electronic level, standing for the lowest available
orbital of the molecule; when an electron resides on the
level, it interacts (linearly) with an Einstein oscillator. Our analysis does not include electronic interactions, but focuses on the electron-vibron ones. Thus, the  Hamiltonian of the
molecular bridge, which includes the coupling with the vibrations,
reads
\begin{align}
{\cal H}^{}_{\rm M}=\epsilon^{}_{0}c^{\dagger}_{0}c^{}_{0}+\omega^{}_{0}( b^{\dagger}b+\frac{1}{2})+\gamma (b+b^{\dagger})c^{\dagger}_{0}c^{}_{0}\ , \label{HD}
\end{align}
where $\epsilon^{}_{0}$ is the energy of the localized level,
$\omega^{}_{0}$ is the frequency of the harmonic oscillator
representing the vibrations, and $\gamma$ is its coupling to the
electrons. The Hamiltonian describing the tunneling
between   the molecule and the leads is
\begin{align}
{\cal H}^{}_{\rm coup}=\sum_{k}(V^{}_{k}c^{\dagger}_{k}c^{}_{0}+{\rm Hc})+\sum_{p}(V^{}_{p}c^{\dagger}_{p}c^{}_{0}+{\rm Hc})\  \label{HCL}
\end{align}
[using $k(p)$ for the left (right) lead].
The leads' Hamiltonian is
\begin{align}
{\cal H}^{}_{\rm lead}={\cal H}^{}_{\rm L}+{\cal H}^{}_{\rm R}+{\cal H}^{}_{\rm LR}\ .
\end{align}
Here
\begin{align}
{\cal H}^{}_{\rm L(R)}=&\sum_{k(p)}\epsilon^{}_{k(p)}c^{\dagger}_{k(p)}c^{}_{k(p)}
\  \label{HL}
\end{align}
is the Hamiltonian of each of the leads
and
\begin{align}
{\cal H}^{}_{\rm LR}=\sum_{kp}V^{}_{kp}e^{i\phi}c^{\dagger}_{k}c^{}_{\rm p}+{\rm Hc}\ \label{HDC}
\end{align}
describes the direct coupling between the two leads (pictorially
shown as the lower arm of the ring in Fig. \ref{sys}). The A-B magnetic flux is included in the 
phase factor $\exp[i\phi]$, see Eq. (\ref{phi}).   %, the flux $\Phi$ is measured in units of
%the flux quantum, $hc/e$. 
Since we use units in which $\hbar =1$,
the flux quantum is $2\pi c/e$. Thus, our model Hamiltonian is
\begin{align}
{\cal H}={\cal H}_{\rm lead}+{\cal H}^{}_{\rm M}+{\cal H}^{}_{\rm coup}\ ,\label{FH}
\end{align}
where the operators $c^{\dagger}_{0}$, $c^{\dagger}_{k}$, and
$c^{\dagger}_{p}$ ($c^{}_{0}$, $c^{}_{k}$, and $c^{}_{p}$) create
(annihilate) an electron on the level, on the left lead, and on
the right lead, respectively, while $b^{\dagger}$ ($b$) creates
(annihilates) an excitation of frequency $\omega_{0}$ on the
molecule.

In the spirit of the Landauer approach,  \cite{book} the various
reservoirs  (which are assumed in the simplest case to be  large 
enough to stay in equilibrium in spite of the small currents that 
flow from/to them) and which supply charge and energy to the leads, 
are described by equilibrium distributions with given chemical potentials 
and temperatures. This keeps the populations on each lead 
of the excitations pertaining to a specific reservoir to be in equilibrium as well. The
electronic reservoirs on the left and on the right of the bridge
are characterized by the electronic distributions  $f^{}_{\rm L}$
and $f^{}_{\rm R}$,
\begin{align}
f^{}_{\rm L(R)}(\omega )=\Bigl (1+\exp [\beta^{}_{\rm L(R)}(\omega
-\mu^{}_{\rm L(R)})]\Bigr )^{-1}\ , \label{FLR}
\end{align}
determined by the respective Fermi functions, with $\beta^{}_{\rm
L(R)}=1/k^{}_{\rm B}T^{}_{\rm L(R)}$. The phonon reservoir, which
determines the vibration population on the bridge, is
characterized by the Bose-Einstein distribution,
\begin{align}
N=\Bigl (\exp[\beta^{}_{\rm V}\omega^{}_{0}]-1\Bigr )^{-1}\ ,\label{NDEF}
\end{align}
with $\beta^{}_{\rm V}=1/k^{}_{\rm B}T^{}_{\rm V}$.

\section{A-B oscillations in the transmission of the interferometer and their decoherence} \label{sec5}
%Here we use the notation of the previous section. 

We  start by considering  the interferometer having just
a resonance level without vibrational coupling on one arm and a direct transmission on the other. The interacting with the vibrations is later included.

\subsection{A simple interferometer: no vibronic coupling}
\noindent
The transport through the ``direct" arm of the interferometer (the one not carrying the dot) is characterized by the dimensionless coupling
\begin{align}
\lambda (\omega )=(2\pi)^{2}\sum_{k,p}|V^{}_{kp}|^{2}\delta (\omega -\epsilon^{}_{k})\delta (\omega -\epsilon^{}_{p})\ ,
\end{align}
such that the bare transmission (reflection) amplitude through that branch is $t_{o}$ ($r_{o}$),
\begin{align}
t^{2}_{o}(\omega )=\frac{4\lambda (\omega )}{[1+\lambda (\omega )]^{2}}\ ,\ \ \ r^{2}_{o}=1-t^{2}_{o}\ .\label{t0}
\end{align}
The transport through the arm carrying the quantum dot is characterized by
\begin{align}
\Gamma^{}_{\rm L(R)}(\omega )=\pi \sum_{k(p)}|V^{}_{k(p)}|^{2}\delta (\omega -\epsilon^{}_{k(p)})\ .
\end{align}
Below we denote the total width of the level $\epsilon_{0}$ of the quantum dot by $\Gamma (\omega )$,
with
\begin{align}
\Gamma (\omega )=\frac{\Gamma^{}_{\rm L}(\omega )+\Gamma^{}_{\rm R}(\omega )}{1+\lambda (\omega )}\ ,
\end{align}
and the asymmetry of the coupling of the dot to the leads by $\alpha (\omega )$,
\begin{align}
\alpha^{2}(\omega )=\frac{4\Gamma^{}_{\rm L}(\omega )\Gamma^{}_{\rm R}(\omega )}{[\Gamma^{}_{\rm L}(\omega )+\Gamma^{}_{\rm R}(\omega )]^{2}}\ .
\label{alfa}\end{align}

\noindent In Ref. ~\cite{Oramnon} it was found that the current, $I$, and the (electronic) energy current, $I_{\rm E}$,  across the interferometer are determined by the same transmission function,
\begin{align}
I=&-e\int\frac{d\omega}{2\pi}[f^{}_{\rm R}(\omega )-f^{}_{\rm L}(\omega )]
{\cal T}(\omega )\ ,\nonumber\\
I^{}_{\rm E}(\omega )&=
\int\frac{\omega^{} d\omega}{2\pi}[f^{}_{\rm R}(\omega )-f^{}_{\rm L}(\omega )]
{\cal T}(\omega )\ ,\label{IIECUR}
\end{align}
where the Fermi functions in the leads, $f_{\rm L,R}$ are given in Eq. (\ref{FLR}).
%\begin{align}
%f^{}_{\rm L,R}(\omega )=[e^{\beta^{}_{\rm L,R}(\omega -\mu^{}_{\rm L,R})}+1]^{-1}\    .
%\end{align}
The function ${\cal T}(\omega )$ is  \cite{Oramnon} % [Eq. (38) of Ref. ~\onlinecite{Hofstetter,Oramnon}]
\begin{align}
{\cal T}_{}(\omega )&=t^{2}_{o}\Bigl (1-\Gamma{\rm Im}G^{a}_{}+\frac{\Gamma^{2}}{4}[1-\alpha^{2}_{}\cos^{2}\Phi ]|G^{a}_{}|^{2}\Bigr )\nonumber\\
&+t^{}_{o}r^{}_{o}\Gamma\alpha\cos\Phi{\rm Re}G^{a}_{}
+\frac{\Gamma^{2}\alpha^{2}}{4}|G^{a}_{}|^{2}\ .\label{ELTR}
\end{align}
Here we have omitted the explicit $\omega-$dependence of (some of) the functions for brevity.
The first term in this expression, $t^{2}_{o}$,  yields the conductance in the
absence of the ring arm carrying the bridge (see Fig. \ref{sys}),
while the last term yields the conductance of that arm alone. The
other three terms in Eq. (\ref{ELTR}) result mainly from various
interference processes (see below) .

The advanced Green function of the dot is
\begin{align}
G^{a}_{}(\omega )=\frac{1}{\omega -\widetilde{\epsilon}^{}_{0}(\omega )-i\Gamma(\omega )/2}\ ,
\end{align}
where $\widetilde{\epsilon}_{0}(\omega )$ is the resonance energy shifted by the coupling to the other parts of the interferometer,
\begin{align}
\widetilde{\epsilon}^{}_{}(\omega )=\epsilon^{}_{0}+\frac{1}{1+\lambda (\omega )}[\lambda (\omega )\Gamma^{}_{\rm L}(\omega )\Gamma^{}_{\rm R}(\omega )]^{1/2}_{}\cos\phi\ .
\end{align}
\noindent
 This result describes well  many
interferometer experiments. It satisfies the Onsager-related symmetry of the conductance, which follows from the  property that ${\cal T}(\omega )$
is even in the flux $\Phi$.

We consider for simplicity the  symmetric case, $\alpha^2(\omega ) = 1$,
which will be assumed in the rest of this section;
all $V$'s are taken to be real (time-reversal symmetry at zero flux).
% and take $|t_0| << 1$. Neglecting  $O(|t_0|^2)$ terms, eq.~\ref{11} is then approximated by
%\begin{align}
%{\cal T}(\omega )&=
%|G^{r}_{}(\omega )|^{2}[
%t^{}_{o}r^{}_{o}\Gamma (\omega -\widetilde{\epsilon}^{}_{0})\cos\phi\nonumber\\
%& +\frac{\Gamma^{2}}{4}]\ .\label{appr}
%\end{align}
%\noindent
%It is seen that this is very similar to just the absolute value squared of the sum of the direct transmission amplitude
%and the one through the branch with the resonant level.}
We now turn to the explanation of the three terms in Eq.  (\ref{ELTR}) which are not direct transmissions through one of the arms. The second term in the brackets is just a correction to the direct transmission $t_0$ due to attempting to go through the bridge (and instead being reflected) either before or after the direct transition. The third term is the  ``weak localization"
correction to the transmission, due to the increase of the total reflection via the two time-reversed paths which encircle
the whole ring clockwise and anti-clockwise.
The term before the last, having the product of $t_0$ and (real part of) the transmission amplitude through the bridge, $\Gamma{\rm Re}G^{a}_{}$, resembles the mixed term in the good old two-wave
interference. There is however a very important small difference: the latter should have the real part of
$\Gamma G^{a}_{}  e^{i\phi}$, while our expression has $\cos\phi {\rm Re}G^{a}_{}$. It thus has
a pure $\cos \phi$ dependence on the flux as it should. This is so,  because we consider a {\it closed}
interferometer which {\it conserves particles}. This differs from the ordinary two-wave interference, in which particles are lost
and therefore, a $\sin(\phi)$ dependence is also generated in the transmission (``no phase rigidity",  see Ref. ~ \cite{AS}). Technically, this is due to the fact that here the Green's function of the dot, which is determined by  paths starting at and returning to it,  is dressed  by paths (or ``diagrams") in which the electron experiences the whole interferometer. Physically, the point is that the fact that electrons can enter and exit the interferometer from/to the two leads, does {\em not} break particle conservation for the scattering matrix of the interferometer. Therefore, the Onsager symmetry, for which particle conservation, or``unitarity", is essential, \cite{AS,LL} holds.

%In the next subsection, we introduce the coupling with the local vibrations, %where, as expected, the $cos(\phi)$
%term disappears from the transmission when the electron exchanges energy with the vibration 
%and discuss how $O(\cos^2(\phi)$ terms may and do appear.

%I see no reason not to add to this (the essence of) Holstein, namely dress $t_0$ with an
%$O(\gamma^2)$ process where the same phonon  is virtually absorbed going to the dot
%and reemitted going from it to the 2nd lead. When this phonon will be in "resonance"
%with the (renormalized) dot level, we'll get the delta-fcn and the i, paving our way to
%fame and wealth.. Because of Onsie, the sinphi can only appear in off-diagonal transport.

\subsection{Interferometer with vibronic coupling}
% The  Hamiltonian of the
%localized state, which includes the coupling with the local vibration,
%reads
%\begin{align}
%{\cal H}^{}_{\rm M}=\epsilon^{}_{0}c^{\dagger}_{0}c^{}_{0}+\omega^{}_{0}( b^{\dagger}b+\frac{1}{2})+\gamma (b+b^{\dagger})c^{\dagger}_{0}c^{}_{0}\ , \label{HD}
%\end{align}
%where $\epsilon^{}_{0}$ is the energy of the localized level,
%$\omega^{}_{0}$ is the frequency of the harmonic oscillator
%representing the vibrations, and $\gamma$ is its (linear) coupling to the
%transport  electrons. The coupling occurs when the electron is on the localized level.

The technical calculations of the transport coefficients with the electron-vibron coupling $\gamma$ have been done in Ref.~\cite{Oramnon}.
They are rather complicated, and so far it has only been feasible to do them to order $\gamma^2$, which is sufficient to see the effects of vibronic excitation/deexcitation. The full results to this order are given in  that reference %Ref. ~\cite{Oramnon} 
and there is no point to repeat them here. Suffice it to say that, for the electrical  conductance, indeed the occurrence of a vibronic change of state does eliminate, in the term resulting from the interaction with vibrations,  the $\cos \phi$ term in the conductance, as expected due to decoherence. It was however found that
a $\cos^2 \phi$ (or $\cos 2 \phi$) dependence survives. This dependence resembles the weak-localization-type flux dependence, as in the
third term of Eq. (\ref{ELTR}) discussed above. This in fact results from a novel generalization of the usual weak-localization process, in which the two time-reversed paths change the vibrational state {\em in the same fashion} \cite{Stern}. Thus, there is no decoherence of this special process!
The situation with  the thermal and thermoelectric coefficients is much more complicated and interesting. It will be reviewed in the rest of this paper.

\section{Generalities on thermoelectric transport}\label{sec6}
The electronic part of thermoelectric linear-transport problem is fully characterized for the two-terminal situation by
\begin{align}
\left( \begin{array}{c}
I^{}_e\\ I_Q^e\end{array}\right) =
\left( \begin{array}{cccc}
G & L^{}_1  \\
L^{}_1 & K_e^0  \\
\end{array}\right) \left(\begin{array}{c}
\delta\mu/e\\ \delta T/T \end{array}\right) \ ,\label{FIRO}
\end{align}
where $I_e$ is the charge current and $I_Q$ the heat current,  $\delta T = T_l - T_r$ and $\delta \mu /e \equiv V$ is the voltage between the left and right terminals. The $2\times 2$ matrix contains the regular conductance $G$, the bare electronic thermal conductance, $K_e^0$ and the (off-diagonal) thermoelectric coefficients $L_1$. That the latter two are equal is the celebrated Onsager relation (valid for time-reversal symmetric systems; with a magnetic field, $B$, if one of them is taken at $B$, it is equal to the other one at $-B$). We remind the reader that since the (Seebeck) thermopower is the voltage developed due to a temperature difference at zero electrical current,  $S = L_1 / G$.
All currents and transport coefficients in Eq. (\ref{FIRO}) are given in 1D for noninteracting electrons in terms of integrals involving the energy-dependent elastic transmission, ${\cal T}(E)$ and the Fermi function, $f(E)$, at the common chemical potential $\mu$ and temperature $T$, 
$G = I_0; L^{}_1 = I_1; K_e^{0} = I_2$, where

\begin{align}
 I_n= \frac{2e}{h}\int_{-\infty}^\infty dE {\cal T}(E)  f(E)[1-f(E)](E-\mu)^n/(k_{\rm B}^{}T).\label{I}
\end{align}
We recall that since the heat conductance is defined as the heat current due to a unit temperature difference, {\em at vanishing electrical current}, the electronic heat conductance is given by
\begin{align}
K_e^{} = K_e^0 - GS^2  = I_2 - I_1^{2}/ I_0\label{K}\ .\end{align}
In a thermoelectric  energy conversion device, 
%(such as a thermal engine or a thermoelectric cooler), it turns out that its 
the performance is governed by the well-known dimensionless   ``figure of merit", $zT$,
\begin{align}
zT = \frac{L^2}{K G} = \frac{GS^2}{K},  \label{fom}
\end{align}
where $K = K_e +K_{ph}$ is the full thermal conductance. Here $K_{ph}$ is the phonon ($+$ any other neutral mobile excitation of the solid) thermal conductance.
The efficiency is a fraction, $g(zT)$, of the, maximal allowed, Carnot efficiency.  
Large $zT$'s yield better performance, with $g(zT)$ rising monotonically from $0$ for $zT \rightarrow 0$ to $1$ for $zT \rightarrow \infty$.

An important remark here is that the determinant of the transport matrix in Eq. (\ref{FIRO}) is nonnegative. 
%When $K_{ph}$ is added to $K_e^0$ this becomes stronger. 
This implies that were the figure of merit based on the bare conductance $K^0 = K_e^0 +K_{ph}$, $zT$ would be limited by unity. It is the ``renormalization"  of $K$ as in Eq. (\ref{K}), which allows larger $zT$ values and opens the way to thermoelectric applications!

Based on the above, Mahan and Sofo,  \cite{MAHAN} in an important development, suggested that regarding ${\cal T}(E) f(E)[1-f(E]/G$ as a (positive, normalized) weight function,
then $S= <E-\mu>$ and $K_e^{} = <(E-\mu)^2> - <E-\mu>^2$. One can then make $K_e^{}$
very small by having a very narrow transmission band away from the Fermi level.
This is needed in order to have a finite $<E-\mu>$, not relying only on the, usually small, asymmetry of 
${\cal T}(E)$ near $E_{\rm F}$ , to break electron-hole symmetry. This implies that $zT$ would be limited only by $K_{ph}$.
Tricks for reducing $K_{ph}$  by acoustic and geometrical mismatch, have been suggested. \cite{Dresselhaus} 
%A significant feature of such a setup  is that the electronic heat
%conductance can be made to be very small  while the thermopower stays finite.

As an example we take an impenetrable barrier of height $W>>k_{\rm B}T$, whose transmission we take as ${\cal T}(E)\simeq \Theta(E-W)$. We measure all energies from the common chemical potential $\mu$. Under these conditions, 
$f(E)$   to the Boltzmann distribution, leading to $f(E)[1-f(E)]\simeq \exp[-E/(k^{}_{\rm B}T)]$. Hence, 
\begin{align} 
<E> = eL^{}_{1}/G&=W+k^{}_{\rm B}T\ ,\nonumber\\
%\frac{eL_1}{G} = k_BT(x_w +1)&\\
<E^2>-<E>^2 &= (k_{\rm B}^{}T)^2\ ,
\end{align}%\exp(-\frac{E}{k_BT})$. We have
%\begin{align} 
%<E> = eL^{}_{1}/G&=k^{}_{\rm B}T(x^{}_{w}+1)\ ,\nonumber\\
%\frac{eL_1}{G} = k_BT(x_w +1)&\\
%$<E^2>-<E>^2 &= (k_{\rm B}^{}T)^2\ ,
%\end{align}
with the charge and heat electron current given by Eqs. (\ref{IIECUR}).
%\begin{align}
%I &= \frac{2e}{h}\int_0^\infty dE [f_L(E)-f_R(E)]{\cal T}(E)\ ,\nonumber\\  
%I_Q &= \frac{2}{h}\int_0^\infty dE [f_L(E)-f_R(E)]E{\cal T} (E)\ .
%\end{align}
%where $x_{w}=W/k^{}_{\rm B}T$.
To reiterate,  the formulae above say that $S$ is the average energy transferred by an electron,  divided by $e$,  while $K_e^0/G$ is the average of the square of that energy, divided by $e^{2}$. Therefore $K_e/G$ is proportional to the variance of that energy.  The latter obviously vanishes for a very narrow transmission band. In the high barrier  case that   band is a few $k_{\rm B} T$ above W. Thus, it is not surprising that $K_e$ is of the order of $(k_{\rm B} T / W)^2$.

\section {Interference effects and their decoherence in thermoelectric transport} \label{sec7}

The linear-transport full set of thermoelectric coefficients \cite{Oramnon}
is given by
\begin{align}
\left [\begin{array}{c}I\\ I^{}_{\rm Q} \\-\dot{E}^{}_{\rm V}\end{array}\right ]
={\cal M}
\left [\begin{array}{c}\Delta\mu/e\\  \Delta T/T \\ \Delta T^{}_{\rm V}/T\end{array}\right ]\ ,\label{MAT}
\end{align}
where the matrix of the transport coefficients, ${\cal M}$, is
\begin{align}
{\cal M}=\left [\begin{array}{ccc}{\rm G}(\Phi )\ \ & \ \  {\rm K} (\Phi )\ \ & \ \ {\rm X}^{\rm V}_{}(\Phi )\\
 {\rm K}(-\Phi )\ \ &\ {\rm K}^{}_{2}(\Phi )\  & \ \ \widetilde{\rm X}^{\rm V}_{}(\Phi )\\
 {\rm X}^{\rm V}(-\Phi )\ \  & \ \ \widetilde{\rm X}^{\rm V}_{}(-\Phi )\ \ &\ \ {\rm C}_{}^{\rm V}(\Phi )\end{array}\right ]\ .
\label{M}
\end{align}
This matrix satisfies the Onsager-Casimir relations. All  three diagonal entries are even functions of the flux
$\Phi$. The off-diagonal entries of ${\cal M}$ consist each of a
term even in the flux, and another one, odd in it, obeying
altogether ${\cal M}_{ij}(\Phi ) ={\cal M}_{ji}(-\Phi )$.

The flux dependence of the transport coefficients is very
interesting. We have found (see Ref. ~\cite{Oramnon} for
details) that there are three types of flux dependencies hiding in
those six coefficients. First, there is the one caused by
interference. The interference processes modify the self energy of
the Green functions pertaining to the A-B ring, in
particular the broadening of the electronic resonance level due to
the coupling with the leads. The interference leads to terms
involving $\cos\phi$. Secondly, there is the flux dependence which
appears in the form of $\cos 2\phi$ (or alternatively,
$\sin^{2}\phi$). This reflects the contributions of time-reversed
paths. These two  dependencies are even in the flux. They yield
the full  flux dependence of the diagonal entries of the matrix
${\cal M}$, Eq. (\ref{M}), and the even (in the flux) parts of the
off-diagonal elements. Finally, there is the {\em odd} dependence
in the flux, that appears as $\sin\phi$. This dependence
characterizes the odd parts of the off-diagonal entries of ${\cal
M}$. These terms {\it necessitate} the coupling of the electrons to
the vibrational modes. They arise from Holstein-type processes,
briefly discussed toward the end  of section \ref{dec}.

\section{Examples and discussion}\label{sec8}
As is clear from the results the previous section and  from
Refs. ~\cite{NEWWE} and ~\cite{Oramnon},  the interesting effect induced exclusively  by the
coupling of the electrons to the vibrational modes is the
possibility to create an electric current, or an electronic heat
current,  by applying a temperature difference $\Delta T^{}_{\rm
V}$ on the phonon bath thermalizing this mode. These new
thermoelectric phenomena are specified by the two coefficients
 ${\rm X}^{\rm V}$ and  $\widetilde{\rm X}^{\rm V}$, see Eq.   (\ref{M}),
respectively.  All other 
transport coefficients related to the electronic currents are
mainly due to the transport of the electrons between the
electronic terminals, with slight modifications from the (small)
coupling to the vibrations. We therefore confine the main
discussion in this section to the coefficients
 ${\rm X}^{\rm V}$ and  $\widetilde{\rm X}^{\rm V}$. To make a closer
 connection with possible experiments, we introduce the (dimensionless)  coefficients
\begin{align}
{\rm S}^{\rm V}_{}=e\beta\frac{ {\rm X}^{\rm V}_{}}{\rm G}\ ,
\end{align}
and
\begin{align}
\widetilde{S}^{\rm V}_{}=\frac{\widetilde {\rm X}^{\rm V}_{}}{\rm K_{2}^{}}\ .
\end{align}
The first gives the potential drop across the molecular bridge
created by $\Delta T^{}_{\rm V}$ when the temperature drop there,
$\Delta T$, vanishes, and the second yields the temperature
difference created by $\Delta T^{}_{\rm V}$ when $\Delta\mu=0$
[or the inverse processes, see Eqs. (\ref{MAT}) and (\ref{M})].
For both the conductance, G, and the thermal conductance, K$_{2}$,
we use below their leading terms, resulting from the coupling to
the leads  alone. % (the numerators result from the coupling to the vibrations, and hence are already of order $\gamma^2$).

As  mentioned above, the transport coefficients of our
three-terminal junction obey the Onsager-Casimir relations. They
do it  though in a somewhat unique way:  the ``off-diagonal"
elements are related to one another by the reversal of the
magnetic field. However, they are not a purely odd functions of it.  A
special situation arises when the molecule is connected {\em
symmetrically} to the two leads. In that case, the asymmetry
parameter %$\overline{\alpha}$ vanishes, while 
$\alpha =1$ [see
Eq. (\ref{alfa})]. % the auxilliary parameter $\overline{\alpha}$, 
%defined in Ref. ~\onlinecite{Oramnon} vanishes]. 
Then, the two coefficients,
${\rm X}^{\rm V}$ and  $\widetilde{\rm X}^{\rm V}$,  are odd
functions of the flux. 
%(resembling the thermal Hall-effect
%discussed recently \cite{LEE} in connection with quantum magnets)
\begin{align}
{\rm X}^{\rm V}_{}(\Phi )=e\omega^{}_{0}\sin \phi
\int\frac{d\omega }{\pi} {\cal T}^{\rm V}_{}(\omega ,\Phi )\Bigl
(t^{}_{o}(\omega^{}_{-})-t^{}_{o}(\omega^{}_{+})\Bigr )\
,\label{XP1}
\end{align}
and
\begin{align}
&\widetilde{\rm X}^{\rm V}_{}(\Phi )=\omega^{}_{0}\sin \phi \int\frac{d\omega}{\pi}{\cal  T}^{\rm V}_{}(\omega ,\Phi )\nonumber\\
&\times\Bigl (t^{}_{o}(\omega ^{}_{-})(\omega^{}_{-}-\mu )-t^{}_{o}(\omega ^{}_{+})(\omega^{}_{+}-\mu)\Bigr )\ .
\label{XPT1}
\end{align}
Here, $\omega_{\pm}=\omega+\pm\omega^{}_{0}/2$, and all other quantities are given in %The frequencies $\omega_{+,-}$ are defined in Eq. (35) and the quantity ${\cal  T}^{\rm V}_{}(\omega ,\Phi )$ is defined 
in Eqs. (30), (43), and (54) of Ref. ~\cite{Oramnon}.
In other words, the thermoelectric processes described by ${\rm
X}^{\rm V}_{}(\Phi )$ and $\widetilde{\rm X}^{\rm V}_{}(\Phi )$
require a certain symmetry-breaking. In the absence of the
magnetic field, that is supplied by the spatial asymmetry of the
junction; in the presence of a flux, those processes appear also
for a  junction symmetrically coupled to the leads, provided that the couplings
to the leads depend on the energy.

When the two leads connected to the electronic reservoirs are identical (making the molecular bridge symmetric) the
 transmission amplitude of the direct bond between the two leads, $t^{}_{o}$ [see Eq. (\ref{t0}], is an even function of $\omega $. The transmission
function ${\cal T}^{\rm V}(\omega ,\Phi )$ given in Ref.  ~\cite{Oramnon}, is not entirely even or odd in
$\omega $, and therefore {\it a priori} the integrals which give
${\rm X}^{\rm V}_{}(\Phi )$ and $\widetilde{\rm X}^{\rm V}_{}(\Phi
)$ do not vanish. However, the asymmetry in the
$\omega-$dependence of the integrand (which results from the
$\omega-$dependence of the Green function) is not significant. As
a result, ${\rm S}^{\rm V}$ is extremely small, while
$\widetilde{\rm S}^{\rm V}$ is not (because of the extra $\omega$
factor in the integrand), see Fig. \ref{SYM}. These plots are
computed using $\Gamma(\omega )=\Gamma^0\sqrt{1-(\omega /W)^{2}}$,
and $\lambda (\omega )=\lambda^0 [1-(\omega /W)^{2}]$, where $W$
is half the bandwidth, and all energies are measured in units of
$(1/\beta = k^{}_{\rm B}T)$ (we have set $\Gamma^0=\lambda^0 =1$
and $W=50$).

\begin{figure}[ hbtp]
\includegraphics[width=6.cm]{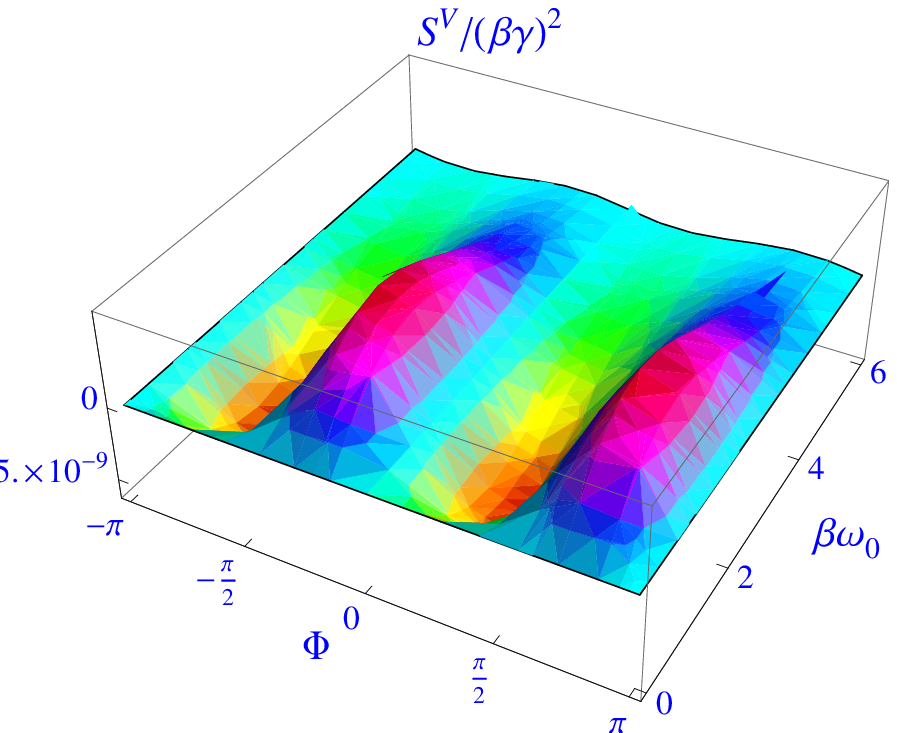}
\includegraphics[width=6cm]{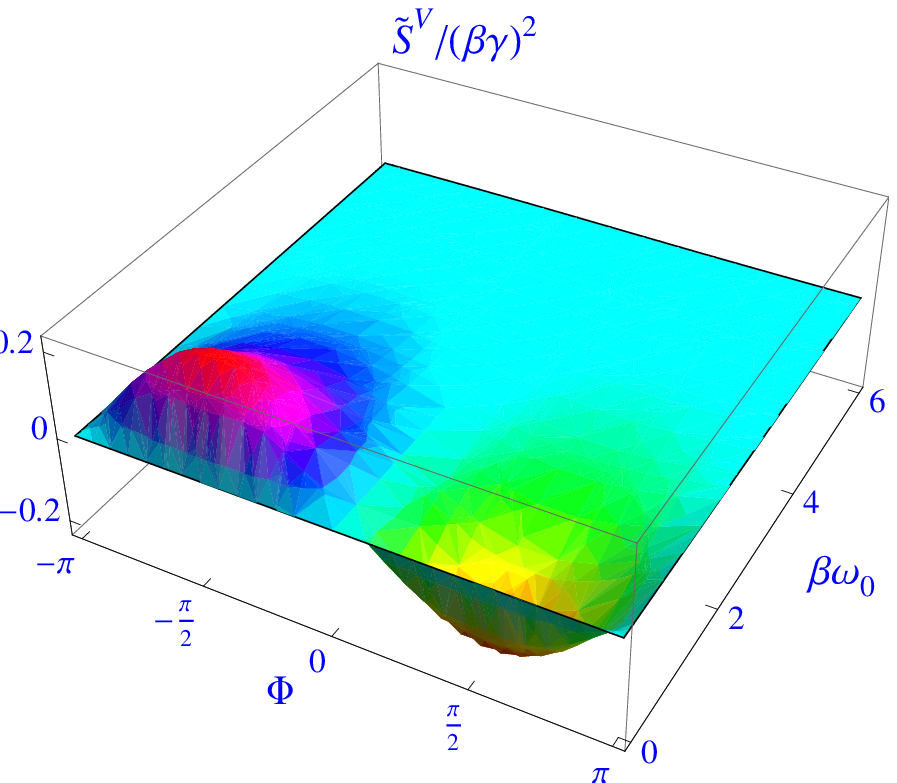}
\caption{(color online.) The transport coefficients ${\rm S}^{\rm V}$ and $\widetilde{\rm S}^{\rm V}$ as functions of the flux (measured in units of the flux quantum) and $\beta\omega^{}_{0}$, for a symmetric bridge.
 } \label{Fig4}
\end{figure}

The relative magnitudes of ${\rm S}^{\rm V}$ and
$\widetilde{S}^{\rm V}$ are significantly changed when the
molecule is coupled {\em asymmetrically} to the leads. Let us  
assume that the left reservoir is
represented by an electron band, such that the partial width it
causes to the resonant level is given by
\begin{align}
\Gamma^{}_{\rm L}(\omega )=\Gamma^{0}_{\rm L}\sqrt{\frac{\omega
-\omega^{}_{c}}{\omega^{}_{v}-\omega^{}_{c}}}\ ,\label{GAMLW}
\end{align}
while the right reservoir is modeled by a hole band, with
\begin{align}
\Gamma^{}_{\rm R}(\omega )=\Gamma^{0}_{\rm
R}\sqrt{\frac{\omega_{v}
-\omega^{}_{}}{\omega^{}_{v}-\omega^{}_{c}}}\ .\label{GAMRW}
\end{align}
The corresponding quantity pertaining to the lower arm of the ring in Fig. \ref{sys} is
\begin{align}
\lambda (\omega )=\lambda^0\sqrt{\frac{\omega_{v}
-\omega^{}_{}}{\omega^{}_{v}-\omega^{}_{c}}}\sqrt{\frac{\omega
-\omega^{}_{c}}{\omega^{}_{v}-\omega^{}_{c}}}\ .\label{GAMDW}
\end{align}
Here, $\omega^{}_{c}$ is the bottom of the conductance band (on
the left side of the junction), while $\omega^{}_{v}$ is the top
of the hole band (on the right one).  The energy integration
determining the various transport coefficients is therefore
limited to the region $\omega^{}_{c}\leq\omega\leq\omega^{}_{v}$.
(For convenience, we normalize the $\Gamma$'s by the full band
width, $\omega^{}_{v}-\omega^{}_{c}$.) Note that the density of states 
is increasing (decreasing) with energy in the electron (hole) lead. Exemplifying  
results in
such a case are shown in Fig. \ref{NONSYM}, computed with
$\Gamma^0_{\rm L}=\Gamma^0_{\rm R}=\lambda^0 =1$ [see Eqs.
(\ref{GAMLW}), (\ref{GAMRW}), and (\ref{GAMDW})], and
$\omega^{}_{c}=-\omega^{}_{v}=100$, all in units of $k_{B}T$.

\begin{figure}[ hbtp]
\includegraphics[width=6.cm]{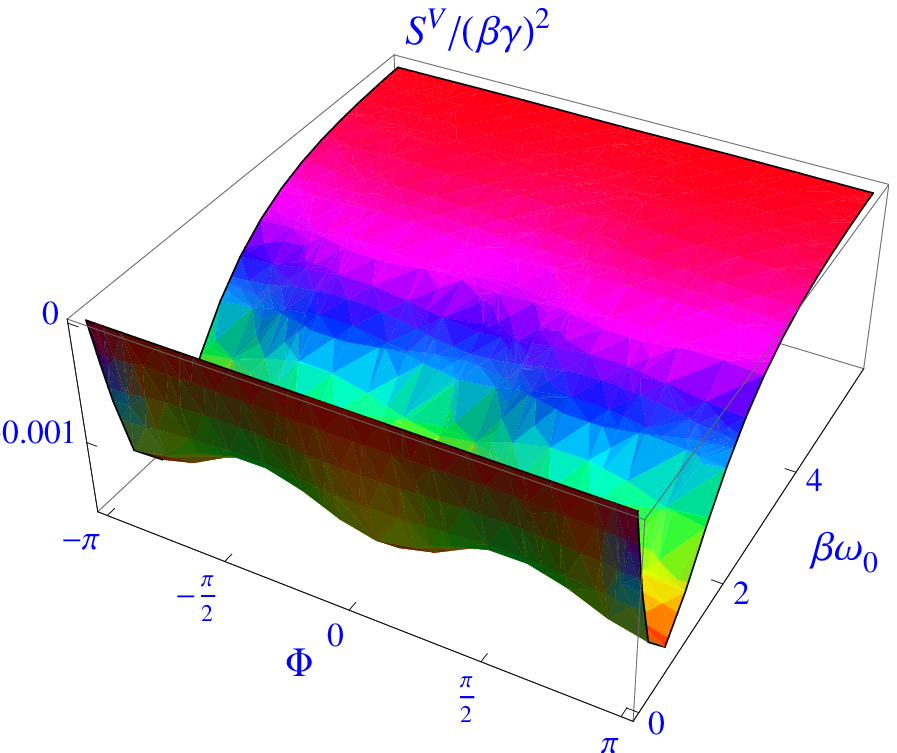}
\includegraphics[width=6cm]{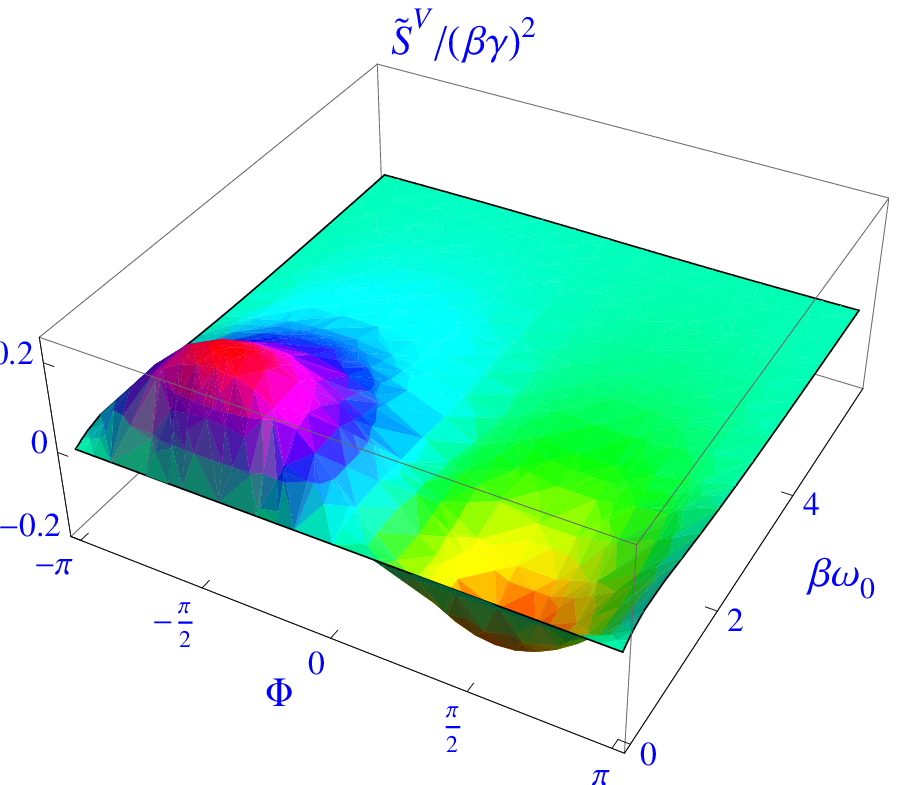}
\caption{(color online.) Same as Fig. 4, for an asymmetric bridge.} 
\label{Fig5}
\end{figure}

Remarkably enough, the coefficient $\widetilde{\rm S}^{\rm V}$,
which measures the ability of the device to turn a temperature
difference on the phonon bath thermalizing the molecular
vibrations into a temperature difference across the molecule is
not very sensitive to the details of the model, apparently because
the term proportional to $\sin\phi$ in Eq. (66) of Ref. ~\cite{Oramnon} is the
dominant one. This means that by reversing the direction of the
magnetic field, one reverses the sign of the temperature
difference or alternatively, the sign of the electronic heat
current. On the other hand, the coefficient ${\rm S}^{\rm V}$,
which sets the scale of the capability to turn a temperature
difference across that phonon bath into an electric current, is
far more sensitive to the details of the model (as expressed e.g.
in our choice for the density of states on the leads) and is less
affected by the magnetic field.

\section{Conclusions}
We reviewed the interference effects of mesoscopic-scale solid-state A-B interferometers,
with special emphasis on how they are reduced/eliminated by decoherence. 
Coupling to local vibrations, to lowest significant order, as well as to a heat bath, was 
introduced. The latter makes the problem much  richer. We first discussed just the electronic transport, with and without the vibronic coupling.
Then, we reviewed the electric and thermal transport in an interferometer
with the above-mentioned couplings. 
In particular, we have found that the thermoelectric transport
coefficients through a vibrating molecular junction, placed on an
Aharonov-Bohm interferometer, have an interesting dependence on
the magnetic flux. The coefficients which relate
the temperature difference between the phonon and electron
reservoirs to the charge and heat currents carried by the
electrons, which exist only due to the electron-vibron coupling,
can be enhanced by the magnetic flux. One of them is also very sensitive 
to the spatial asymmetry.

\section{Acknowledgments}
This work was supported %by the German Federal Ministry of
%Education and Research (BMBF) within the framework of the
%German-Israeli project cooperation (DIP), 
by the US-Israel
Binational Science Foundation (BSF), by the Israel Science
Foundation (ISF) and by its Converging Technologies Program.
OEW and AA are indebted to the Albert Einstein Minerva Center
for Theoretical Physics and to the Goldschleger Center for Nanophysics 
at the Weizmann Institute, for partial support.

\end{document}